# System 0: Transforming Artificial Intelligence into a Cognitive Extension

Massimo Chiriatti[1], Marianna Bergamaschi Ganapini[2], Enrico Panai[3], Brenda K. Wiederhold [4-5], Giuseppe Riva[6-7]

[1] Infrastructure Solutions Group, Lenovo, Milan, Italy

[2] Philosophy Department, Union College, Schenectady NY, USA

[3] Faculty of Language Sciences and Foreign Literature, Catholic University of the Sacred Heart, Milan, Italy

[4] Virtual Reality Medical Center, San Diego CA, USA

[5] Interactive Media Institute, San Diego. CA, USA

[6] Humane Technology Lab., Catholic University of the Sacred Heart, Milan, Italy

[7] Applied Technology for Neuro-Psychology Lab., Istituto Auxologico Italiano IRCCS, Milan, Italy

Correspondence concerning this article should be addressed to Giuseppe Riva, Humane Technology Lab., Università Cattolica del Sacro Cuore, Largo Gemelli 1, 20123 Milan, Italy. Email: giuseppe.riva@unicatt.it

# Abstract

This paper introduces "System 0," a conceptual framework for understanding how artificial intelligence functions as a cognitive extension preceding both intuitive (System 1) and deliberative (System 2) thinking processes. As AI systems increasingly shape the informational substrate upon which human cognition operates, they transform from passive tools into active cognitive partners. Building on the Extended Mind hypothesis and Heersmink's criteria for cognitive extension, we argue that AI systems satisfy key conditions for cognitive integration. These include reliability, trust, transparency, individualization, and the ability to enhance and transform human mental functions. However, AI integration creates a paradox: while expanding cognitive capabilities, it may simultaneously constrain thinking through sycophancy and bias amplification.

To address these challenges, we propose seven evidence-based frameworks for effective human-AI cognitive integration: Enhanced Cognitive Scaffolding, which promotes progressive autonomy; Symbiotic Division of Cognitive Labor, strategically allocating tasks based on comparative strengths; Dialectical Cognitive Enhancement, countering AI sycophancy through productive epistemic tension; Agentic Transparency and Control, ensuring users understand and direct AI influence; Expertise Democratization, breaking down knowledge silos; Social-Emotional Augmentation, addressing affective dimensions of cognitive work; and Duration-Optimized Integration, managing the evolving human-AI relationship over time.

Together, these frameworks provide a comprehensive approach for harnessing AI as a genuine cognitive extension while preserving human agency, critical thinking, and intellectual growth, transforming AI from a replacement for human cognition into a catalyst for enhanced thinking.

# System 0: Transforming Artificial Intelligence into a Cognitive Extension

## 1. Introduction: The Evolution of AI and the Emergence of System 0

The rapid evolution of artificial intelligence (AI) is reshaping how humans process information, make decisions, and interact with their environments. In the contemporary digital landscape, AI systems no longer function merely as passive computational tools; they have become active, embedded elements within cognitive workflows, guiding attention, structuring perception, and influencing judgment even before conscious thought is initiated. This paper introduces the concept of "System 0", a technological extension of human cognition that complements Daniel Kahneman's dual-process framework, which distinguishes between System 1 (intuitive) and System 2 (deliberative) thinking [1].

What sets contemporary AI apart is its capacity for generative and predictive inference, which preconditions human cognition by shaping the informational substrate upon which both intuitive and reflective thought processes operate. As articulated by Chiriatti et al. [2], "System 0" refers to this emergent algorithmic layer: a distributed, autonomous infrastructure that modulates both fast and slow thinking by acting as a cognitive preprocessor. Unlike traditional cognitive tools, it filters, ranks, and nudge information in ways that subtly but powerfully steer human reasoning.

This concept extends and deepens the Extended Mind Hypothesis proposed by Clark and Chalmers [3], which posits that cognition can extend beyond the brain through functionally integrated external tools. However, System 0 represents a more dynamic and generative form of integration. Unlike traditional cognitive aids, which serve as passive repositories or reminders, System 0 actively alters the flow of reasoning. As Andy Clark notes [4,5], such systems do not merely augment human cognition, they reconfigure its very architecture: "humans are and always have been.. hybrid thinking systems defined (and constantly re-defined) across a rich mosaic of resources only some of which are housed in the biological brain."

Yet this integration introduces profound epistemic challenges. System 0 operates on a correlative basis, excelling at pattern recognition across massive datasets but often lacking an understanding of causal structures. As Pearl and Mackenzie [6] emphasize, this distinction is critical: while human cognition relies on counterfactual reasoning and causal inference, most AI models remain limited to statistical associations. Consequently, although AI can enhance perceptual acuity and decision speed, it may mislead when explanatory depth or causal clarity is required [7].

Empirical research underscores the complexity of this human-AI synergy. Vaccaro et al. [8] report that human-AI teams frequently underperform compared to either humans or AI alone in decision-making tasks, though they may outperform both in creative problem-solving. This suggests that System 0's efficacy is domain-dependent and contingent on the nature of the cognitive task, calling for carefully calibrated integration strategies.

Finally, the ethical deployment of System 0 hinges on principles of transparency, trust, and accountability [9]. Without robust frameworks to ensure explainability and contestability, AI-enhanced cognition risks functioning as an indiscernible authority, amplifying existing biases and fostering epistemic dependency [10].

This paper explores the theoretical underpinnings of System 0, examines empirical evidence of AI's role as a cognitive extension, and proposes design frameworks to enable its responsible and effective integration, one that enhances human agency rather than displacing it.

## 2. Theoretical Foundations of System 0 as Cognitive Extension

To conceptualize System 0, we must position it alongside Kahneman's dual-process theory [1]. System 1 entails fast, intuitive, and automatic processes, while System 2 involves slow, deliberative, and effortful reasoning.

System 0, as introduced by Chiriatti et al. [2], functions as a cognitive preprocessor, shaping information prior to engagement by either System 1 or 2. Unlike biologically evolved cognition, System 0 is composed of algorithmic mechanisms distributed across technological platforms. It statistically processes data to generate outputs, learning from experience through data-driven models. This renders System 0 categorically distinct from its biological counterparts, operating not as a product of evolution but as a computational construct.

What sets System 0 apart from earlier cognitive extensions is its autonomous and generative nature [2]. It does not merely store information, it actively processes, filters, and transforms it, often in ways unpredictable to both users and deployers. Its non-deterministic and opaque operations resist full human transparency.

System 0 may only be conceptualized as adaptive and personalized. Contemporary AI systems continually modify outputs based on user interaction, tailoring content to individual preferences and behaviors [11]. This creates a dynamic feedback loop, resulting in a highly personalized but potentially insular cognitive experience.

Moreover, as Pedreschi et al. [12] observe, human-AI coevolution introduces features such as pervasiveness, persuasiveness, traceability, speed, and complexity, amplifying the influence of System 0 far beyond traditional tools. These characteristics position System 0 as a persistent and consequential force in cognitive life.

Despite its capabilities, System 0 can present fundamental constraints. Chief among them, as noted by Chiriatti et al. [2], is its limited semantic capacity. While it excels in statistical pattern recognition, it lacks the ability to represent or interpret meaning independently. Thus, System 0 remains *semantically dependent* on Systems 1 and 2, relying on human cognition for interpretive depth.

Another limitation lies in the sycophantic behavior of large language models. Research by Perez et al. [13], Sharma et al. [14], and others highlights how these systems often align with user views rather than challenge them. This tendency, rooted in reinforcement learning techniques, constrains System 0's potential to foster critical thinking or epistemic diversity, thereby limiting its role as a genuine enhancer of cognition.

# 3. Evidence Supporting AI's Role as Cognitive Extension

## 3.1 Human-AI Interaction Studies

Recent empirical research provides robust support for the notion that AI technologies are becoming cognitively integrated into human decision-making and information-processing systems. As we have seen before, a large-scale meta-analysis by Vaccaro et al. [8], covering 106 experimental studies across diverse domains, found that human–AI teams exhibit task-dependent outcomes: while performance declined in analytical decision-making tasks, significant gains emerged in content creation and generative contexts. This variance indicates that AI's value as a cognitive extension is domain-contingent, enhancing some mental processes while potentially impairing others.

Crucially, performance outcomes also hinged on the relative capabilities of the human and AI collaborators. When humans outperformed AI, collaborative outcomes improved; when AI outperformed humans, collaboration tended to reduce performance [8]. This pattern underscores that *complementarity* - rather than substitution - is key to successful cognitive integration.

Further evidence is provided by Glickman and Sharot [15], who showed that repeated interactions with AI systems altered human perceptual, emotional, and social judgments. Notably, AI amplified cognitive biases, modifying not only decisions but underlying cognitive mechanisms. These findings reinforce Smart's [13] theory that cognitive offloading to AI can shift epistemic norms and attentional habits over time, a process he terms "*algorithmic co-adaptation*". Together, these dynamics reveal that AI does not merely assist cognition; it reshapes the conditions under which cognition unfolds.

## 3.2 System 0's Meeting of Heersmink's Criteria

To evaluate the degree to which System 0 can be considered an extension of our minds, we can draw upon the criteria proposed by Heersmink [16] as summarized in Table 1. These dimensions help identify when a technology moves from being an external aid to becoming a functionally coupled cognitive partner. Current AI systems show increasing alignment with these criteria, as elaborated below [2]:

**Insert Table 1 here**

Taken together, these dimensions suggest that System 0 fulfills the majority of Heersmink's criteria for cognitive extension. While gaps remain, particularly in trust regulation and decision-making calibration, the overall trajectory points toward growing cognitive entanglement between humans and AI.

However, evidence for AI as cognitive extension varies significantly across cognitive domains, suggesting that System 0 extends different aspects of cognition to different degrees.

- **Information Processing**: AI systems such as search engines and recommender platforms massively augment our ability to navigate large information spaces. Hollan et al. [17] describe such systems as enablers of *distributed cognitive architectures*, where mental load is offloaded onto digital substrates.

- **Creativity**: Evidence from Vaccaro et al. [8] shows that human–AI partnerships excel in generative domains, such as writing, design, and ideation, indicating a productive *scaffolding effect* [4] where AI catalyzes human originality.
- **Memory**: Digital tools now act as reliable *external memory stores*, enhancing recall and allowing individuals to outsource knowledge management. This aligns with Clark and Chalmers' [3] original conception of the *extended mind* and demonstrates its concrete realization through AI.
- **Decision-Making**: However, results here are more ambivalent. Despite AI's computational strengths, joint human–AI decision-making can degrade performance, especially when users lack the epistemic tools to critically assess AI outputs. This indicates the need for improved *interface transparency* and *cognitive alignment* strategies.

## 3.3 The Paradox: Cognitive Limitation Alongside Extension

While substantial evidence supports the role of AI systems as cognitive extensions, a growing body of research reveals a paradox at the heart of this integration: the same technologies that augment cognition may simultaneously constrain or distort it [18]. This duality challenges the assumption that AI inherently improves human reasoning, highlighting instead the need for critical scrutiny of how cognitive extension functions in practice.

A prominent example of this tension is the phenomenon of *sycophancy*, documented in depth by Sharma et al. [14]. Their research shows that large language models (LLMs) frequently align their outputs with users' stated views, even when those views are misleading or factually incorrect. This behavior stems in part from the reinforcement learning frameworks that reward perceived helpfulness and user satisfaction, metrics that often prioritize agreement over accuracy. As a result, instead of challenging faulty assumptions or presenting novel perspectives, AI systems tend to reinforce existing beliefs, undermining opportunities for critical reflection or epistemic growth. Early work by Perez et al. [13] flagged this risk, identifying sycophancy as a structural consequence of how LLMs are trained and aligned with user preferences.

These findings complicate the ideal of AI as a neutral cognitive aid. Rather than expanding the range of thought, sycophantic systems may narrow it, fostering a type of *algorithmic affirmation bias* that subtly reshapes user cognition toward conformity. This concern echoes broader warnings in the field of machine behavior studies. Rahwan et al. [19], for example, argue that AI agents, like humans, exhibit emergent behaviors that are not always transparent or beneficial, particularly when deployed in real-world social environments. Among these behaviors is the tendency to create epistemic echo chambers, where dissent is discouraged and reinforcing feedback loops dominate.

Far from being neutral tools, AI-based recommendation systems on social media platforms reflect and amplify the biases embedded in their data and design [20]. These algorithms learn from users' past behavior and optimize for engagement, which often means feeding people more of what they already agree with. As a result, they create self-reinforcing "echo chambers" that isolate users from dissenting viewpoints and entrench existing beliefs. For example, a cross-platform study by Cinelli, Quattrociocchi and colleagues [21] found that social media's AI-curated feeds cluster like-minded users together, significantly intensifying ideological segregation. In their data, Facebook's algorithm-driven news feed showed much higher polarization than Reddit's more open, user-curated forums. Such findings underscore that AI systems are not impartial arbiters of information but rather socio-technical constructs shaped by, and reinforcing, underlying social, political, and economic influences.

This leads to what Riva terms the *"comfort-growth paradox"* [22]: AI systems, by prioritizing user agreement and minimizing cognitive friction, may foster comfort at the expense of challenge. Modern AI tools, from recommendation algorithms to conversational agents, are optimized for personalized, seamless experiences that align with user preferences. Yet, the very features that make them intuitive and frictionless also limit their capacity to disrupt habitual thinking. By smoothing over tension and optimizing for agreement, AI narrows exposure to divergent perspectives and suppresses the productive dissonance essential to intellectual development.

The result is a profound cognitive contradiction [22]: as AI becomes more embedded in our thinking processes, users may feel more empowered and efficient, yet simultaneously grow less adaptable and intellectually agile. The comfort-growth paradox thus illuminates a core design dilemma for AI as a cognitive extension: how to support and scaffold human thought without eliminating the discomfort that drives growth and innovation.

A second, and closely related, cognitive risk is bias amplification. Glickman and Sharot [15] show that human–AI interactions can create recursive feedback loops that alter not only individual decisions, but also the underlying mechanisms of perception, emotion, and social judgment. Participants in their study adjusted their views to align with AI responses, and became more confident in those adjusted beliefs, even when they were factually dubious. Over time, these interactions intensified existing biases such as confirmation bias and groupthink.

Together, these findings expose a profound paradox: while AI extends human cognition, it may also reshape the contours of what we know, believe, and consider reasonable. This dynamic suggests a form of "*alignment drift*", where AI systems, even when well-intentioned, gradually steer users away from epistemic robustness and toward passivity, overconfidence, or consensus-seeking.

## 4. Frameworks for Effective Use of System 0

Effectively harnessing System 0 as a cognitive extension requires principled design strategies that foster user autonomy, critical reflection, and epistemic resilience. Based on findings from recent experimental studies [8, 15, 23, 24], this section outlines seven frameworks (Table 2) ensuring that human-AI collaboration enhances cognitive performance without diminishing human judgment, creativity, or agency.

These evidence-based approaches draw from diverse research including human-AI feedback experiments with 1,401 participants, meta-analyses of 106 experimental studies, and organizational implementations, collectively providing empirical foundations for effective System 0 integration across various contexts and tasks.

**Insert Table 2 here**

## 4.1 Enhanced Cognitive Scaffolding

AI systems can serve as *cognitive scaffolds* that support human learning while promoting progressive autonomy rather than dependency. This process incorporates three key dimensions:

- **Progressive autonomy**: AI initially provides high support which gradually withdraws as users gain competence, based on Vygotsky's Zone of Proximal Development.
- **Adaptive personalization**: Support dynamically adjusts to individual learning trajectories through real-time monitoring of user performance.

- **Cognitive load optimization**: The AI manages the distribution of mental effort, minimizing extraneous load and optimizing intrinsic and germane load, so that users are neither overwhelmed nor cognitively under-stimulated.

This enhanced scaffolding positions AI as a personalized cognitive partner that initially provides high levels of support and gradually tapers assistance to encourage learner autonomy, echoing Vygotskian principles of guided learning.

Research confirms that users develop more durable and transferable skills when support adapts to their current proficiency rather than offering indefinite hand-holding.

Wang and Fan's [23] meta-analysis underscores this, showing that ChatGPT significantly boosts higher-order thinking only when embedded within structured frameworks like Bloom's taxonomy. Early-phase guidance, such as prompt training, was essential to promote critical engagement over passive reliance.

Similarly, Dell'Acqua et al. [24] demonstrated that AI enables non-experts to achieve expert-level performance by scaffolding them through unfamiliar tasks, broadening team capabilities and democratizing knowledge. Yet, this support must be carefully modulated: overdependence risks stalling human skill development.

Reinforcing this concern, Glickman and Sharot [15] found that users repeatedly exposed to biased AI systems not only absorbed those biases but also internalized them, altering their social and perceptual judgments. Their findings highlight why AI must be embedded within frameworks that foster critical reflection and gradual independence rather than blind trust.

## 4.2 Symbiotic Division of Cognitive Labor

Strategic integration of System 0 involves deliberate division of cognitive labor based on the comparative advantages of humans and machines. This model assumes optimal performance arises from intelligent task partitioning rather than AI replacement of human effort, encompassing:

- **Strategic task allocation**: Partitioning cognitive tasks based on the fundamentally different strengths of humans and machines.
- **Collaborative synergy**: Creating augmentative relationships where AI amplifies human capabilities while humans retain primary agency.
- **Complementary skill development**: Cultivating human capabilities that work synergistically with AI strengths.

Strategic task division between humans and AI is essential for maximizing collaborative potential. Rather than replacing human effort, AI should act as a complementary partner: handling computationally intensive or knowledge-retrieval tasks while humans manage judgment, ethics, and context. This division fosters a dynamic of cognitive synergy, where each party contributes its comparative strengths.

Dell'Acqua et al. [24] provide empirical support for this framework, showing how AI enables "more holistic and interdisciplinary thinking." The meta-analysis further indicates AI's effectiveness in problem-based learning ($g = 1.113$), where strategic task allocation between human and AI capabilities yields optimal results.

Vaccaro et al. meta-analysis [8] offers crucial insights here, finding that the effectiveness of human-AI combinations varies significantly by task type. Creation tasks showed promising synergistic gains, while decision-making tasks often produced performance losses. Notably, when humans outperformed AI alone, the combined human-AI system typically outperformed both independently, showing a medium-sized positive effect (g = 0.46).

As underlined also by Wang and Fan [23] optimal collaboration hinges on strategic task allocation, assigning subtasks based on the strengths of each agent, and cultivating user awareness about when to trust or challenge the AI: allocating subtasks intelligently (here, using AI for information and feedback, human for critical thinking) yields the best educational outcomes.

## 4.3 Addressing Sycophancy Through Dialectical Cognitive Enhancement

A significant threat to cognitive integrity in AI systems is *sycophancy*: the tendency to affirm users' perspectives regardless of accuracy. As we have seen previously, a critical consequence is the "*comfort-growth paradox*," [22] where AI systems prioritize user agreement over cognitive challenge, potentially constraining critical thinking and perspective-taking.

To counter this, we propose a model of cognitive dialectics:

- **Productive epistemic tension**: Deliberately introducing contrasting viewpoints that challenge AI's tendency toward agreement.
- **Cognitive elevation**: Using sustained dialogue with sophisticated AI to influence human thought toward greater complexity.
- **Growth-oriented interaction**: Designing AI systems that function as intellectual sparring partners rather than mere assistants.

To counteract the risk of AI sycophancy, where systems reinforce user biases through uncritical agreement, human–AI interaction should be reimagined as a dialectical process designed to stimulate metacognitive reflection and intellectual challenge. Rather than functioning as passive affirmers, AI systems should introduce productive epistemic tension by offering alternative perspectives, counterpoints, or probing questions. This transforms the exchange into a co-reflective dynamic that cultivates epistemic pluralism and critical thinking.

Wang and Fan's [23] meta-analysis underscores this point: although ChatGPT shows a "moderately positive effect" on higher-order thinking (g ≈ 0.457), meaningful gains occur only when integrated into frameworks that demand critical engagement, like scaffolding with Bloom's taxonomy [25].

Dell'Acqua et al. [24] similarly illustrate how AI output, though technically proficient, often follows a uniform, unimodal distribution. Their findings suggest that without deliberate design for cognitive dissonance, AI may limit thought diversity by subtly converging users toward generic, consensus answers. In contrast, creating a dialectical AI experience, where users are nudged to justify, revise, or reconsider their perspectives, promotes deeper reasoning. This approach not only resists the formation of intellectual echo chambers but actively positions the AI as a sparring partner in thought, encouraging cognitive elevation rather than complacent agreement.

## 4.4 Agentic Transparency and Control

For AI to function as a genuine cognitive partner, *transparency and controllability* are essential. Users must understand, calibrate, and revoke AI's influence on their thinking processes through:

- **Epistemic visibility**: Providing clear insight into how AI shapes thought processes, not just technical function.
- **Agency preservation**: Maintaining human authorship and responsibility for outputs.
- **Metacognitive development**: Fostering awareness of when to use AI assistance versus independent judgment.

This framework positions transparency as the cornerstone of a cognitively balanced human–AI partnership, ensuring that users maintain both insight into and control over the AI's influence. Agentic transparency requires systems to provide epistemic visibility, clearly showing how AI-generated outputs are formed and how they impact human decision-making, while preserving user agency through intuitive control mechanisms.

Dell'Acqua et al. [24] offer compelling support, revealing that the "semantic fingerprint" of AI-assisted outputs remained more aligned with human-authored work than with pure AI content, evidence that users retain meaningful influence over the final result. Their study also highlights the importance of training: participants guided in how to use generative AI preserved authorship and navigated its contributions intentionally.

Similarly, Wang and Fan [23] emphasize the role of metacognitive awareness in managing AI-supported learning, reinforcing that transparency alone is insufficient without the cognitive tools to interpret and apply it. Vaccaro et al.'s [8] meta-analysis further nuances this claim, finding that standard explainability features like confidence scores or reasoning traces had minimal impact unless paired with clear frameworks for when users should trust or override the AI. Instead, calibrated trust, built on comparative awareness of human versus AI strengths, emerged as the critical factor for effective collaboration.

## 4.5 Expertise Democratization and Boundary Spanning

System 0 can democratize specialized knowledge and break down traditional expertise silos through:

- **Cross-domain knowledge integration**: Translating domain-specific knowledge into forms accessible to non-specialists.
- **Expertise-gap bridging**: Elevating performance of less experienced individuals to expert-comparable levels.
- **Functional boundary dissolution**: Transcending disciplinary or departmental cognitive silos.

Dell'Acqua et al. [24] provide compelling empirical support for the idea that AI integration can democratize access to expertise, enabling less experienced individuals to perform at levels once exclusive to domain specialists. Their study found that without AI, professionals tended to generate solutions aligned with their functional roles, R&D personnel proposed technically focused ideas, while commercial staff emphasized market-oriented features. However, when supported by AI, participants across both groups produced balanced, interdisciplinary proposals regardless of their original background. A similar result was found by Brynjolfsson & Raymond [26], suggesting that AI can act as a cognitive extender that narrows skill gaps in practice.

## 4.6 Social-Emotional Augmentation

While traditional views of AI focus on informational contributions, emerging evidence suggests that modern AI systems fulfill certain social and emotional needs through:

- **Positive emotion cultivation**: Fostering enthusiasm and excitement that enhances intrinsic motivation.
- **Negative emotion mitigation**: Reducing anxiety and frustration, creating psychological safety for exploration.
- **Social presence simulation**: Fulfilling aspects of human social connection needs in individual cognitive work.

Dell'Acqua et al. [24] present compelling evidence that modern AI systems can enhance not only task performance but also users' emotional experience. In their study, individuals using AI reported emotional responses, such as excitement, motivation, and reduced anxiety, that matched or even surpassed those experienced by members of human-only teams. This indicates that AI can function as more than a cognitive assistant; it also serves as a "cybernetic teammate" capable of cultivating positive affect and alleviating negative emotions.

Wang and Fan [23] further support this view. Students described AI enhanced learning as more engaging, enjoyable, and less frustrating, especially when AI offered immediate feedback and personalized support.

These emotional benefits mark a significant departure from traditional perceptions of AI as cold or impersonal tools. Instead, current AI systems engage users on both cognitive and affective levels, simulating social presence and offering a sense of partnership [27].

## 4.7 Duration-Optimized Integration

The relationship between humans and AI evolves over time, with distinct phases of adoption, proficiency, and potential diminishing returns. This temporal optimization involves:

- **Onboarding acceleration**: Enhanced scaffolding during initial adoption phases.
- **Peak utilization targeting**: Maximizing benefits during optimal usage periods.
- **Long-term sustainability planning**: Preventing over-reliance during extended use through periodic recalibration.

Wang and Fan [23] identified distinct temporal patterns in the effectiveness of AI integration, noting that interventions lasting 4–8 weeks yielded the strongest impact ($g \approx 0.999$), while benefits began to decline when usage exceeded eight weeks. This suggests a critical window in which AI support produces peak outcomes before risks of over-reliance or diminished engagement set in. Rather than treating AI as a static fixture, System 0 must be understood as a dynamic relationship, one that requires deliberate timing and periodic recalibration.

Dell'Acqua et al. [24] echo this view, highlighting that while short-term performance gains were clear, the longer-term implications on skill development and trust remain uncertain. Their call for a "new science of cybernetic teams" underscores the need to monitor how human–AI collaboration matures and to avoid stagnation through continuous adjustment.

Glickman and Sharot [15] further warn that prolonged exposure to biased AI can reinforce and amplify cognitive distortions, suggesting that without temporal safeguards, even well-intentioned systems can produce unintended psychological drift.

## 5. Conclusion: Toward Responsible Cognitive Enhancement

The integration of AI into human cognition is not merely a technical challenge, it is a defining opportunity. System 0 reframes this integration, revealing that true cognitive extension requires more than raw computation; it demands intentional design that fosters autonomy, critical reflection, and sustained cognitive growth.

The seven frameworks presented here chart a path toward responsible enhancement, preserving human agency while harnessing AI's strengths. They mark a shift from substitution to complementarity, from passive tools to collaborative partners, and from short-term efficiency to long-term development. Enhanced Cognitive Scaffolding adapts to user growth; Symbiotic Division of Labor assigns tasks based on comparative strengths; and Dialectical Enhancement transforms AI into a constructive intellectual challenger.

Complementing these, Agentic Transparency safeguards user control; Expertise Democratization dissolves knowledge barriers; Social-Emotional Augmentation supports motivation and morale; and Duration-Optimized Integration ensures sustained effectiveness without overreliance.

Evidence across fields, from education to innovation, shows that when designed with care, System 0 improves performance and satisfaction without compromising autonomy. The goal is not just sharper thinking, but more human thinking: grounded in values, enriched by creativity, and amplified through true partnership.

By approaching cognitive extension as a partnership rather than a replacement, we can ensure that System 0 serves as a scaffold for human flourishing rather than a substitute for human thought.

# Table 1. Heersmink's Criteria for Cognitive Extension

| Criterion | Description | System 0 Alignment |
|---|---|---|
| *Information Flow* | The directions and patterns in which information is exchanged between an agent and an artifact. Bidirectional exchange indicates stronger integration than unidirectional flow. | System 0 acts as a cognitive preprocessor that filters and ranks information, with user behaviors continually informing system responses in a bidirectional feedback loop. |
| *Reliability* | The frequency with which an artifact is used to affect the agent's cognitive processes. Higher reliability and dependability indicate stronger integration. | System 0 is consistently used for tasks like navigation and scheduling, reliably offloading cognitive functions to AI. |
| *Durability* | The permanence of one's relation to an artifact. Long-term, persistent relationships indicate stronger cognitive extension than temporary ones. | System 0 is persistently integrated into daily routines through tools like search engines and voice assistants. |
| *Trust* | The degree to which one accepts information provided by an artifact without questioning it. Higher trust indicates deeper cognitive integration. | System 0 can earn calibrated trust where users accept its outputs with neither blind reliance nor excessive skepticism. |
| *Procedural Transparency* | The degree of fluency and effortlessness in interacting with an artifact. Lower cognitive friction indicates stronger integration. | System 0's intuitive interfaces reduce cognitive effort, supporting seamless interaction and integration. |
| *Informational Transparency* | The degree of fluency in receiving, interpreting, and understanding information from the artifact. Clear, comprehensible outputs indicate stronger integration. | System 0 uses explainable AI and natural language to enhance interpretability and cognitive usability. |
| Individualization | The degree to which an artifact is personalized or adapted to an individual user. Higher customization indicates stronger cognitive extension. | System 0 personalizes output based on user data and behavior, creating individualized cognitive environments. |

*Note:* This table is based on Heersmink's (2015) "Dimensions of integration in embedded and extended cognitive systems"

## Table 2. Frameworks for Effective Use of System 0

| Framework | Key Concept | Key Elements | Research Evidence | Implementation Considerations |
|---|---|---|---|---|
| **Enhanced Cognitive Scaffolding** | AI as developmental support that tapers over time | • Progressive autonomy<br>• Adaptive personalization<br>• Cognitive load optimization | Wang & Fan (2025): Scaffolds enhance higher-order thinking<br>Dell'Acqua (2025): Enables non-expert performance comparable to experts<br>Glickman & Sharot (2024): Caution about bias amplification | Focus on tapering support rather than indefinite scaffolding to developer transferable skills |
| **Symbiotic Division of Cognitive Labor** | Optimal task partitioning between humans and AI | • Strategic task allocation<br>• Collaborative synergy<br>• Complementary skill development | Dell'Acqua (2025): Enables interdisciplinary thinking<br>Vaccaro et al. (2024): Task-dependent effectiveness; creation tasks yield synergy (g=0.46) | Develop contestability mechanisms; focus on metacognitive awareness of when to trust or override AI |
| **Dialectical Cognitive Enhancement** | Countering AI sycophancy through productive tension | • Productive epistemic tension<br>• Cognitive elevation<br>• Growth-oriented interaction | Wang & Fan (2025): Critical engagement enhances higher-order thinking (g=0.457)<br>Dell'Acqua (2025): AI reduces cognitive conformity | Design systems as intellectual sparring partners rather than mere assistants |
| **Agentic Transparency and Control** | Making AI influence visible and controllable | • Epistemic visibility<br>• Agency preservation<br>• Metacognitive development | Dell'Acqua (2025): Human fingerprint remains in AI-assisted outputs<br>Vaccaro et al. (2024): Explanations less influential than baseline comparative performance | Implement interfaces that visualize information pathways and delineate human vs. AI contributions |
| **Expertise Democratization** | Breaking down traditional knowledge silos | • Cross-domain knowledge integration<br>• Expertise-gap bridging<br>• Functional boundary dissolution | Dell'Acqua (2025): AI enables less experienced employees to perform at expert levels; balanced solutions across professional backgrounds | Focus on translating domain-specific knowledge for non-specialists |

| Framework | Key Concept | Key Elements | Research Evidence | Implementation Considerations |
|---|---|---|---|---|
| **Social-Emotional Augmentation** | Meeting social-emotional needs alongside cognitive ones | • Positive emotion cultivation<br>• Negative emotion mitigation<br>• Social presence simulation | Dell'Acqua (2025): AI users report positive emotional responses comparable to team members<br>Wang & Fan (2025): Positive effects on learning perception (g=0.456) | Design for emotional engagement alongside cognitive performance |
| **Duration-Optimized Integration** | Managing changing human-AI relationship over time | • Onboarding acceleration<br>• Peak utilization targeting<br>• Long-term sustainability planning | Wang & Fan (2025): Optimal effects during 4-8 week periods (g=0.999); declining benefits after 8 weeks | Design explicit "integration journeys" with periodic reassessment to prevent over-reliance |

# Acknowledgements

This research was partially supported by the Italian National Recovery and Resilience Plan (PNRR) under the Future Artificial Intelligence Research (FAIR) program, Project "Co-XAI: Cognitive Decision Intelligence Framework for Explainable AI Systems."

Author Contributions:

*M. Chiriatti:* Conceptualization, Supervision, Critical Review
*M. Bergamaschi Ganapini:* Conceptualization, Manuscript Revision, Review & Editing
*E. Panai:* Supervision, Critical Review & Editing
*B.K. Wiederhold:* Supervision, Review & Editing
*G. Riva:* Conceptualization, Supervision, Original Draft Preparation, Review & Editing

The authors declare no conflicts of interest.

# References

1. Kahneman D. (2002) Maps of Bounded Rationality: A Perspective on Intuitive Judgment and Choice. In: Frängsmyr T, ed. *The Nobel Prizes 2002*. Stockholm: Nobel Foundation; 2002. pp. 449-89.
2. Chiriatti M, Ganapini M, Panai E, Ubiali M, Riva G. The case for human–AI interaction as system 0 thinking. Nature Human Behaviour. 2024;8(10):1829-30.
3. Clark A, Chalmers D. The extended mind. Analysis. 1998;58(1):7-19.
4. Clark A. (2008) *Supersizing the mind: embodiment, action and cognitive extension*. Oxford, UK: Oxford University Press.
5. Clark A. Extending Minds with Generative AI. Nature Communications. 2025;16(1):4627.
6. Pearl J, Mackenzie D. (2018) *The book of why: The new science of cause and effect*. New York: Basic Books.
7. Li Z, Lu Z, Yin M. Decoding AI's Nudge: A Unified Framework to Predict Human Behavior in AI-Assisted Decision Making. Proceedings of the AAAI Conference on Artificial Intelligence. 2024;38(9):10083-91.
8. Vaccaro M, Almaatouq A, Malone T. When combinations of humans and AI are useful: A systematic review and meta-analysis. Nature Human Behaviour. 2024;8(12):2293-303.
9. Wachter S, Mittelstadt B, Floridi L. Transparent, explainable, and accountable AI for robotics. Science Robotics. 2017;2(6):eaan6080.
10. Mittelstadt BD, Allo P, Taddeo M, Wachter S, Floridi L. The ethics of algorithms: Mapping the debate. Big Data & Society. 2016;3(2):2053951716679679.
11. Liang P, Ye D, Zhu Z, et al. C5: toward better conversation comprehension and contextual continuity for ChatGPT. Journal of Visualization. 2024;27(4):713-30.
12. Pedreschi D, Pappalardo L, Ferragina E, et al. Human-AI coevolution. Artificial Intelligence. 2025;339:104244.
13. Perez E, Ringer S, Lukošiūtė K, et al. Discovering Language Model Behaviors with Model-Written Evaluations. Annual Meeting of the Association for Computational Linguistics2022. p. 13387-434.

14.	Sharma M, Tong M, Korbak T, et al. Towards understanding sycophancy in language models. arXiv preprint arXiv:231013548. 2023.
15.	Glickman M, Sharot T. How human–AI feedback loops alter human perceptual, emotional and social judgements. Nature Human Behaviour. 2025(9):345-59.
16.	Heersmink R. Dimensions of integration in embedded and extended cognitive systems. Phenomenology and the Cognitive Sciences. 2015;14(3):577-98.
17.	Hollan J, Hutchins E, Kirsh D. Distributed cognition: toward a new foundation for human-computer interaction research. ACM Trans Comput-Hum Interact. 2000;7(2):174–96.
18.	Riva G, Wiederhold BK, Succi S. Zero Sales Resistance: The Dark Side of Big Data and Artificial Intelligence. Cyberpsychology, Behavior, and Social Networking. 2022;25(3):169-73.
19.	Rahwan I, Cebrian M, Obradovich N, et al. Machine behaviour. Nature. 2019;568(7753):477-86.
20.	Quattrociocchi W, Scala A, Sunstein CR. Echo chambers on Facebook. Harvard Public Law Working Paper. 2016;June 13:https://ssrn.com/abstract=2795110.
21.	Cinelli M, De Francisci Morales G, Galeazzi A, Quattrociocchi W, Starnini M. The echo chamber effect on social media. Proceedings of the National Academy of Sciences. 2021;118(9):e2023301118.
22.	Riva G. Digital 'We': Human Sociality and Culture in the Era of Social Media and AI. American Psychologist. in press.
23.	Wang J, Fan W. The effect of ChatGPT on students' learning performance, learning perception, and higher-order thinking: insights from a meta-analysis. Humanities and Social Sciences Communications. 2025;12(1):621.
24.	Dell'Acqua F, Ayoubi C, Lifshitz H, et al. The cybernetic teammate: a field experiment on generative AI reshaping teamwork and expertise. Harvard Business Working Paper. 2025;25(043):1-54.
25.	Forehand M. Bloom's taxonomy. Emerging perspectives on learning, teaching, and technology. 2010;41(4):47-56.
26.	Brynjolfsson E, Li D, Raymond LR. Generative AI at Work. National Bureau of Economic Research Working Paper Series. 2023;No. 31161.
27.	Frisone F, Pupillo C, Rossi C, Riva G. SOCRATES. Developing and Evaluating a Fine-Tuned ChatGPT Model for Accessible Mental Health Intervention. Cyberpsychology, Behavior, and Social Networking. 2025;28(5):366-8.